\documentclass[A4paper,12pt]{article}
\usepackage[T1]{fontenc}
\usepackage[english]{babel}
\usepackage[cp1250]{inputenc}
\usepackage{epsfig}
\usepackage{amsfonts}
\usepackage{fancyhdr}
\usepackage{syntonly}
\usepackage{graphicx}

\begin{document}
\selectlanguage{english}

\begin{titlepage}
\begin{center}
\vspace*{3cm}

\begin{title}
\bold {\Huge Note on the CMS multiplicity distributions
 }
\end{title}

\vspace{2cm}

\begin{author}
\Large K. FIA{\L}KOWSKI\footnote{e-mail address:
fialkowski@th.if.uj.edu.pl}, R. WIT\footnote{e-mail address:
romuald.wit@uj.edu.pl}

\end{author}

\vspace{1cm}

{\sl M. Smoluchowski Institute of Physics\\ Jagellonian University \\

30-059 Krak{\'o}w, ul.Reymonta 4, Poland}

\vspace{2cm}

\begin{abstract}
The CMS data on the multiplicity distributions are compared with the
results from the 8.142 version of the PYTHIA MC event generator with
the default tuning and realistic impact parameter profile. The
agreement is reasonable. However, it seems to be difficult to
improve further this agreement by tuning  the parameters determining
the multiple parton interactions.
\end{abstract}

\end{center}

\vspace{2cm}



\end{titlepage}

\section{Introduction}

In a recent paper \cite{KFRW2} we have discussed the data on the
multiplicity distributions at the LHC energies. We have confirmed
our earlier observation \cite{KFRW} that the observed fast increase
of the average multiplicity with energy \cite{ALICE2}, \cite{ALICE},
\cite{CMS} follows from the standard parametrizations used in the
PYTHIA 8 event generator \cite{SMS}, \cite{SMS2}. The average
multiplicity as well as the scaled moments of the distribution in
the wide energy range are qualitatively compatible with these
parametrizations.
\par
In this note we discuss in more detail the CMS data \cite{CMS} on
the multiplicity distributions at three energies and two
pseudorapidity intervals. Introducing a more realistic impact
parameter profile we find a reasonable quantitative agreement of the
data with PYTHIA 8.142 for the default values of its parameters. We
consider also  some further tuning.

\section{Data and model results}

The CMS data contain 132, 12 and 442 thousand events at $0.9$,
$2.36$ and $7$ TeV, respectively. The trigger is selected to remove
the bulk of single diffractive events and the additional corrections
are made to obtain a possibly pure non-single-diffractive (NSD)
sample. The results for the average multiplicity $\overline n$ and
the three lowest scaled moments of the multiplicity distributions
$$c_q = \frac {\overline{n^q}}{\overline{n}^q}$$
for $q=2,3,4$ and for two choices of the central pseudorapidity bin
widths: $\Delta \eta <1$ and $\Delta \eta <2.8$ are shown in Figs.
1-4.

\begin{figure}[h!]
\centerline{ \epsfig{figure=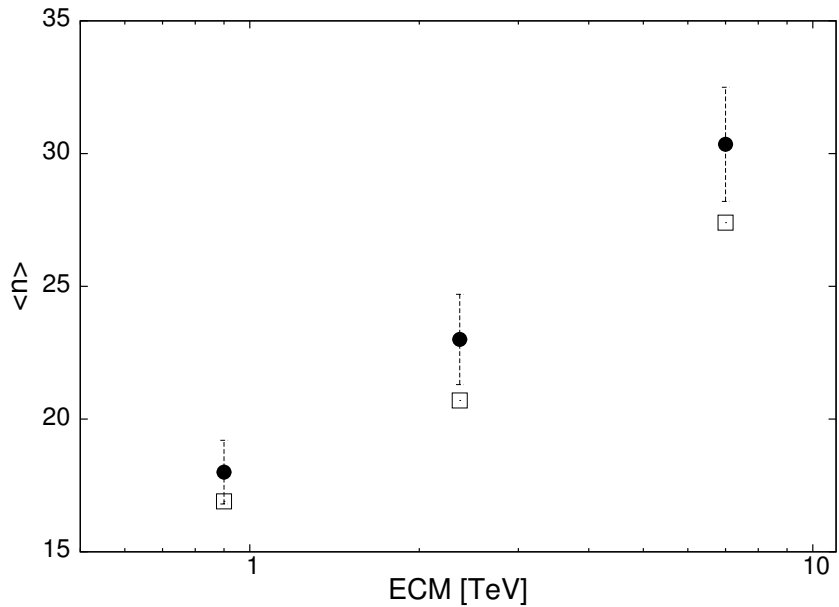,height=8.0cm}}
\caption{\footnotesize \label{Nav} The average multiplicity from the
CMS data at {\it 0.9}, {\it 2.36} and {\it 7} TeV (black dots with
error bars) and from PYTHIA 8.142 (open squares) for the
pseudorapidity bin width of {\it 2.8} are shown.}
\end{figure}

\begin{figure}[h!]
\centerline{ \epsfig{figure=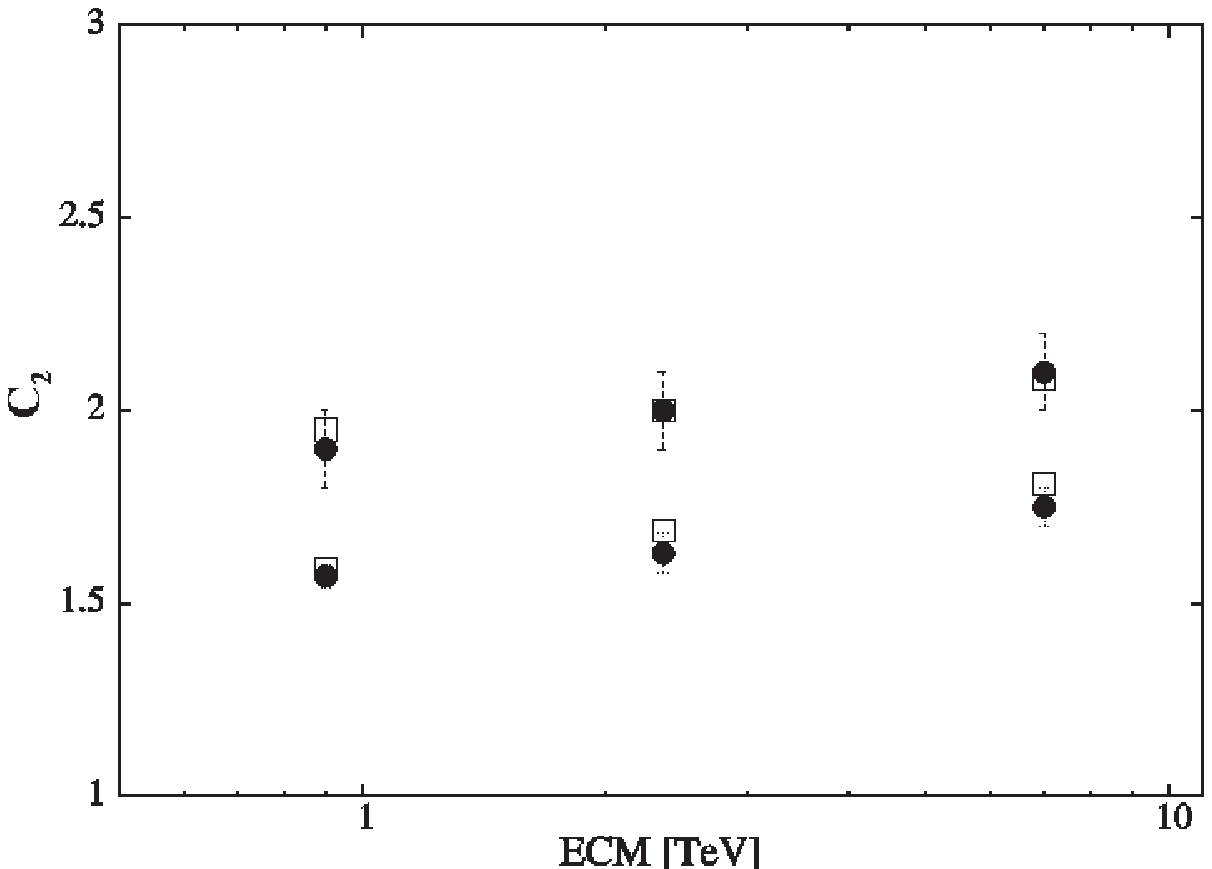,height=8.0cm}}
\caption{\footnotesize \label{C2}The $C_2$ moment from the CMS data
at {\it 0.9}, {\it 2.36} and {\it 7} TeV (black dots with error
bars) and from PYTHIA 8.142 (open squares) for the pseudorapidity
bin widths of {\it 1.0} and {\it 2.8} are shown.}
\end{figure}

\begin{figure}[h!]
\centerline{ \epsfig{figure=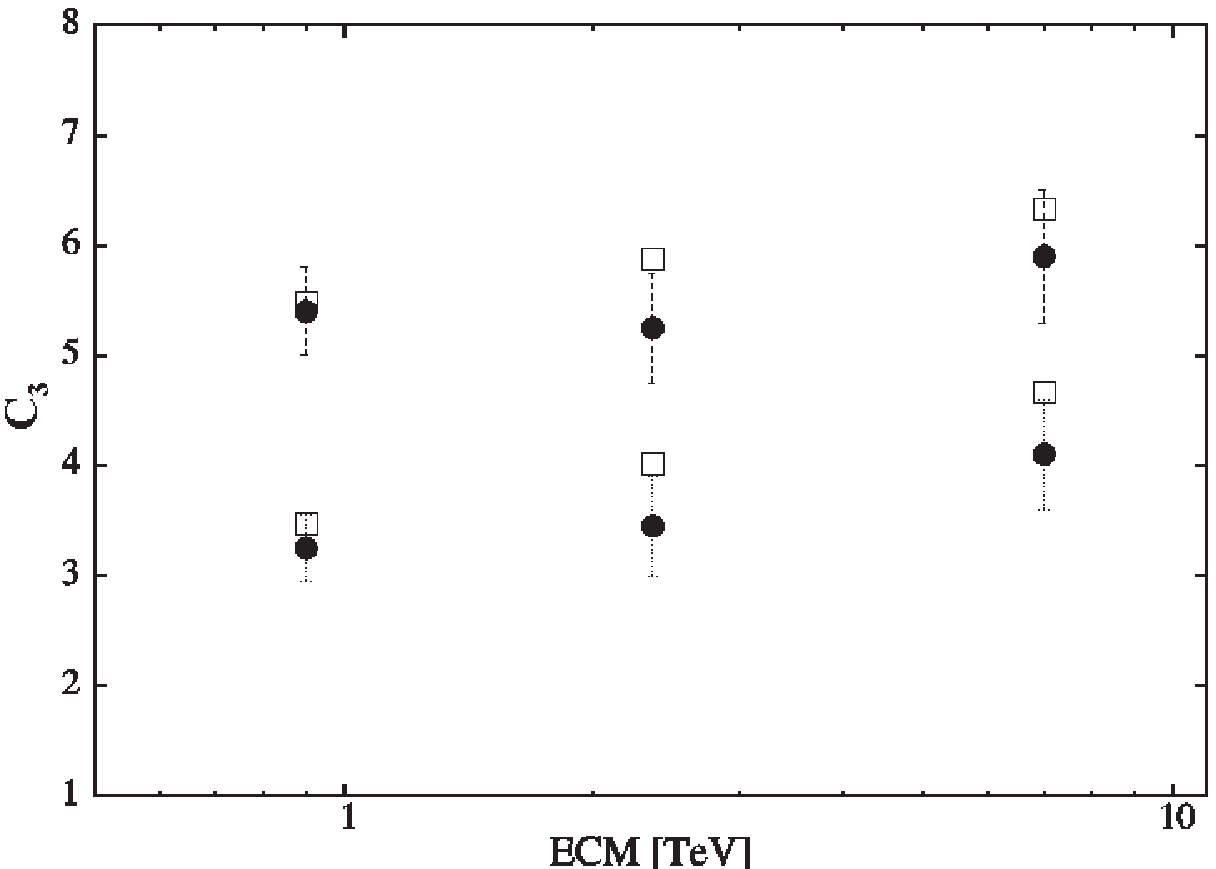,height=8.0cm}}
\caption{\footnotesize \label{C2}The $C_3$ moment from the CMS data
at {\it 0.9}, {\it 2.36} and {\it 7} TeV (black dots with error
bars) and from PYTHIA 8.142 (open squares) for the pseudorapidity
bin widths of {\it 1.0} and {\it 2.8} are shown.}
\end{figure}

\begin{figure}[h!]
\centerline{ \epsfig{figure=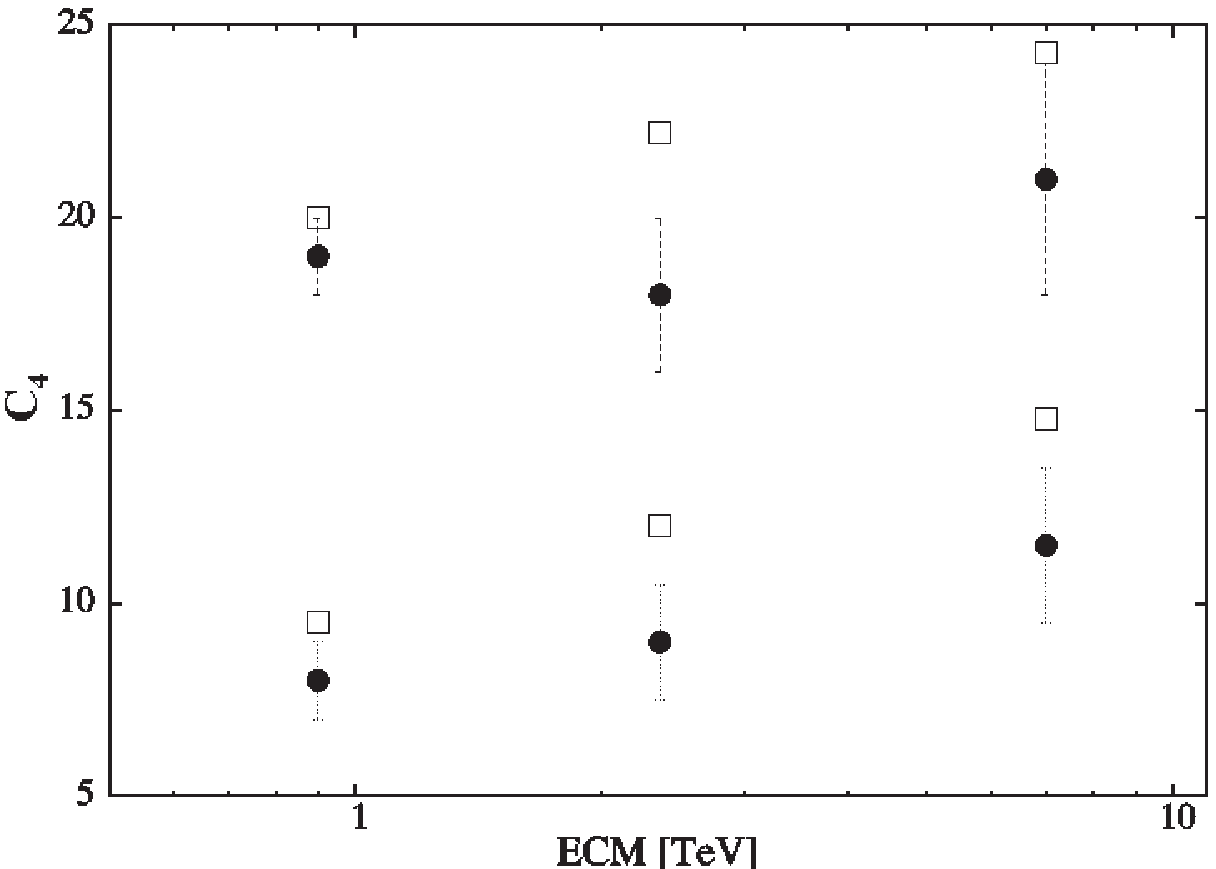,height=8.0cm}}
\caption{\footnotesize \label{C2}The $C_4$ moment from the CMS data
at {\it 0.9}, {\it 2.36} and {\it 7} TeV (black dots with error
bars) and from PYTHIA 8.142 (open squares) for the pseudorapidity
bin widths of {\it 1.0} and {\it 2.8} are shown.}
\end{figure}

The PYTHIA 8.142 is run for NSD events with the default values of
parameters and one modification: the impact parameter profile is not
a single Gaussian, as in the default version, but a (more realistic)
combination of two Gaussians, as used in the earlier versions of
PYTHIA 8. The results are also shown in Figs. 1-4. We repeat that
the PYTHIA parameters are not fitted to these data but are taken
from the default version. Thus this agreement may be regarded as a
good one.
\par
In the previous paper we have compared two different tunings of
PYTHIA 8.142: the default in the 8.135 version. They were bracketing
the experimental results and the differences were growing rather
fast with energy (for the energies below LHC the differences were
negligible). This was due to the different energy dependence of the
lower cut for the transfer momenta used to calculate the multiple
parton interactions. Now we have found that the default version,
modified only in the shape of the b-profile, agrees quite well with
the data and no further tuning seems necessary.

\section{Conclusions and outlook}
We have found a reasonable agreement between the predictions of
PYTHIA 8.142 and the CMS data. However, the detailed comparison of
the model and data shows that the average multiplicity is slightly
underestimated, but he scaled moments are (also slightly)
overestimated. The energy dependence is similarly slightly too weak
for the average but too fast for the scaled moments.
\par We have
checked that tuning the parameters determining the multiple
interactions we increase or decrease simultaneously \textit{both}
the average and the scaled moments. Thus to improve further the
agreement with data it would be probably necessary to tune the
parameters different from that ones we considered here.

\end{document}